\documentclass[a4paper,11pt]{article}
\usepackage{jinstpub} 
\usepackage{subcaption} 
\usepackage{lineno}
\usepackage{filecontents}

\title{\boldmath Application of muon absorption tomography in imaging of civil structures}

\author[a]{Piyush Pallav}
\author[a]{, Purba Bhattacharya}
\author[b,c]{, Supratik Mukhopadhyay}
\author[b]{, Nayana Majumdar}

\affiliation[a]{Department of Physics, Adamas University, Barasat, Kolkata}
\affiliation[b]{Applied Nuclear Physics Division, Saha Institute of Nuclear Physics, Kolkata, A CI of Homi Bhaba National Institute}
\affiliation[c]{Research Wing, Naihati Prolife, Naihati}

\emailAdd{purba.bhattacharya1985@gmail.com}

\abstract{The research focuses on the non-invasive imaging technique using cosmic muon absorption tomography to monitor the internals of archaeological / civil / industrial structures of intermediate size. It integrates experimental measurements and numerical simulations with Geant4, ascertaining the reliability and precision of muon absorption tomography using easily available components for the stated purpose. The experiment probes muon interactions across a range of materials including those commonly used in building civil and industrial structures. An experiment, fondly named MARS (Muon Absorption in Rigid Structures), was carried out to explore the possibility of using overlapped scintillation paddles for improved mapping of inhomogeneities in structures made of concrete. Good correlation of experimental and simulated results for all tests indicates that this simple approach can be  implemented for non-destructive evaluations of structures of civil and industrial interest.}


\begin{document}
\nocite{*}
\maketitle
\flushbottom

\section{Introduction}
\label{sec:intro}

Cosmic ray muon tomography, or muography, is an exciting new technique that uses cosmic rays to give us a view inside objects of archaeological, civil and industrial interest \cite{Sebastien2018}. Essentially, it works by using muons — tiny, high-energy particles created by cosmic rays—to see through these materials. What makes muons special is that they carry around 10,000 times more energy than regular X-rays, which means they can penetrate deeper \cite{Kudryavtsev2021}. One of the best things about this method is that it can help us differentiate between materials and spot hidden issues that traditional methods might completely miss. That said, like any emerging technology, there are still some hurdles - resolution issues and data complexity being two of the big ones \cite{Barnes2023}.

Muography is broadly classified into muon absorption tomography (MAT) and muon scattering tomography (MST) [\cite{Vanini2018, Wen2023, Saracino2018}].  Usually MAT is used for large objects due to which significant amount of muon is absorbed within the object, and it is difficult to place detectors on its either side. MST, on the other hand, needs to identify the amount of scattering of muon tracks by putting detectors on both incoming and outgoing sides of the object. For MST, number of muons getting absorbed by the object is negligible, and the absorption information is ignored altogether.

In this work, we have initiated exploration of the possibility of using a hybrid approach for intermediate-sized structures for which both MAT and MST may be used to improve resulting imaging. By intermediate-sized object we mean objects that are large enough to absorb good number of cosmic muons while they traverse the material, and small enough such that detectors can be placed on either side of it. The absorption data, as well as the scattering data, are obtained from suitably placed detectors around the object. Here, we investigate the possibility of using easily available instruments to implement the muon absorption part of the imaging.

We have worked with a setup called the Cosmic Hunter \cite{Cosmic_Hunter_CAEN}, a commercially available system from CAEN, to carry out the experiments. The idea was to see how well this simple setup, primarily oriented towards educational training, could detect muons and how it could be used to estimate absorption of muons in different materials and imaging of some small structure \cite{Lagrange2023}.  To make sure our experimental results were valid, we compared the measured data with Geant4 \cite{Agostinelli2003, Allison2016} simulations at almost all stages. Based on the studies, we proposed an imaging technique in which overlapped detector plates are used to image a given structure. Although it was challenging to identify muon absorption in concrete (due to its low Z and density), it was possible to identify features of the structure by the overlapped scintillator paddles. This part of the experiment has been christened as MARS (Muon Absorption in Rigid Structures) to indicate that the proposed simple approach can be extended to monitor rigid structures of civil and industrial interest.

In the following sections, we delve deeper into the experimental methodology and findings of our study. The following section \ref{section: setup} describes the experimental setup and methods, focusing on the characterization of the Cosmic Hunter detector, calculations of detector efficiency and muon flux measurements. It also includes details on secondary particle detection. In parallel, the implementation of Geant4 simulation using Cosmic ray shower library (CRY) \cite{CRY}, as event generator is also described. Then in further section \ref{section: absorption} we present the muon absorption study, exploring how muons interact with different materials like aluminum, stainless steel, lead and concrete. In section \ref{section: MARS}, the MARS (Muon Absorption in Rigorous Structures) experiment is introduced where attempts are made to image internals of a given structure using muon absorption information obtained from the simple experimental setup described and validated earlier. Finally, in section \ref{section: conclusion} we summarize the findings of the studies carried out in the context of this report and outline the future plans.

\section{Experimental setup and methods}
\label{section: setup}

The CAEN Cosmic Hunter SP5620CH is a highly sophisticated and flexible system for detecting cosmic rays, and it has been central to our experimental efforts. It is a setup that includes three plastic scintillating tiles and a coincidence module, as shown in Fig.\ref{fig: cosmic hunter}. The detection system uses tiles made from Polystyrene-based material (UPS-923A), also called scintillating tiles. This material has all the key properties needed for effective cosmic ray detection: high light output, great transparency, and fast decay time \cite{cosmic_hunter_manual}. Each scintillator is paired with a Silicon Photomultiplier (SiPM), a photo-detector based on “P-on-N” silicon technology, with a size of 4 x 4 mm² that operates in the Geiger-Müller mode. The SiPM stands out because of its excellent temporal resolution — it can detect light pulses with a rise time of less than 1 nanosecond and a pulse duration of about 2.5 nanoseconds. To ensure the best possible performance, optical grease has been used to enhance the connection between the scintillator and SiPM, and the entire setup is enclosed in black cardboard to block out any ambient light that might interfere with detection.

The coincidence module, which acts as the control center and data acquisition unit, plays a crucial role in the Cosmic Hunter system. It is equipped with an ESP32-based micro-controller (DOIT ESP32 DevKit v1), which allows the user to select events for counting. Researchers can decide whether to count signals from individual scintillating tiles or coincidence signals from multiple tiles. The module has a simple interface, including an e-book display and control buttons, as well as LED indicators that show detection events for each scintillating tile. The user can adjust the integration time for data collection anywhere from 10 minutes to 24 hours. Plus, all the data is automatically recorded on a microSD card in .csv format, The data includes experimental parameters, signals from the scintillators, and even environmental conditions like altitude, temperature, humidity and pressure \cite{cosmic_hunter_manual}. To make sure our setup was reliable, we performed various benchmark experiments under controlled conditions. Our results matched those reported by the manufacturer closely, which gave us confidence in the reliability and accuracy of the system for more advanced experiments involving cosmic rays.

\subsection{Characterization of Cosmic Hunter detector setup}

To assess the performance of the Cosmic Hunter detector setup, a comprehensive characterization was performed. This involved calculating detector efficiency (relative), and measuring the muon flux inside and outside the experimental room and further comparing the muon flux from the results that were predicted by the CRY event generator. 

\subsubsection{Calculation of detector relative efficiency}

\begin{figure}[h]
    \centering
    \includegraphics[width=0.6\textwidth]{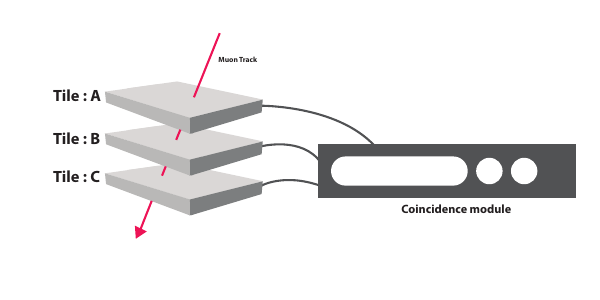}
    \caption{Cosmic hunter muon detections system setup for efficiency measurement.}
    \label{fig: cosmic hunter}
\end{figure}

The relative efficiency of each tile in the Cosmic Hunter setup was calculated by determining the likelihood of muons being detected within the sensitive volume of each scintillating tile. The diagram below (\ref{fig: cosmic hunter}) shows the scintillating tiles and the coincidence module for efficiency measurement.

For three tiles \( A \), \( B \), and \( C \), the calculated relative efficiencies were found to be \( \eta_A = 0.910 \), \( \eta_B = 0.892 \), and \( \eta_C = 0.926 \). The gross relative efficiency of the system \( \eta_{ABC} \) can thus be calculated as:

\[
\eta_{ABC} = \eta_A \cdot \eta_B \cdot \eta_C = 0.910 \times 0.892 \times 0.926 = 0.743
\]

Table~\ref{tab:efficiency_values} presents both the calculated and reference relative efficiency values for each tile. The reference relative efficiencies were obtained from the Cosmic Hunter manual \cite{cosmic_hunter_manual} for comparison, ensuring that relative efficiency of our setup aligns closely with expected values.

\begin{table}[h]
    \centering
    \caption{Calculated and reference relative efficiencies of tiles A, B, and C in the Cosmic Hunter Setup}
    \begin{tabular}{|c|c|c|}
        \hline
        \textbf{Tile} & \textbf{Calculated efficiency} & \textbf{Reference efficiency} \\
        \hline
        A & 0.910(80) & 0.915 \\
        B & 0.892(83) & 0.895 \\
        C & 0.926(88) & 0.930 \\
        \hline
    \end{tabular}
    \label{tab:efficiency_values}
\end{table}

\subsubsection{Cosmic ray shower library (CRY) simulations}

To estimate muon flux accurately, CRY simulations were conducted using input parameters appropriate for the laboratory location. The simulations were performed at sea level with parameters set to an altitude of 0 meters and a latitude of 22.74 degrees corresponding to the experimental site. The total simulation time for 1000 muons over an area of 225 ${cm}^2$ was calculated as follows:

\[
\text{Simulation Time} = 98769.8 \, \text{seconds} = 1646.1 \, \text{minutes}
\]

Based on this simulation time, the muon flux \( \Phi_{\text{CRY}} \) was derived as:

\[
\Phi_{\text{CRY}} = \frac{1000}{1646.1} \approx 0.60 \, \text{muons/cm}^2/\text{minute}
\]

This simulation output provides a benchmark for validating experimental measurements obtained from the Cosmic Hunter setup.

\subsubsection{Muon flux measurements inside and outside the experimental room}

Muon flux measurements were conducted both outside and inside the experimental room to account for the influence of structural shielding. The experimental setup outside the room used three stacked detector plates in ABC coincidence mode, covering an area of 225 cm². The experimentally determined flux outside was:

\[
\Phi_{\text{outside}} = \frac{136.339}{225} \approx 0.61 \, \text{muons/cm}^2/\text{minute}
\]

Measurements inside the room accounted for structural shielding due to concrete layers above the ground floor. The muon flux inside was measured as follows:

\[
\Phi_{\text{inside}} = \frac{111.616}{225} \approx 0.496 \, \text{muons/cm}^2/\text{minute}
\]

For ease of subsequent analyses, this value was approximated to \( 0.5 \, \text{muons/cm}^2/\text{minute} \). The close agreement between simulated and measured flux values confirms the accuracy of the Cosmic Hunter setup in detecting muon flux under varying conditions.

\subsubsection{Detection of Secondary Particles}

\begin{figure}[h!]
    \centering
    \begin{subfigure}[b]{0.6\textwidth}
        \includegraphics[width=\textwidth]{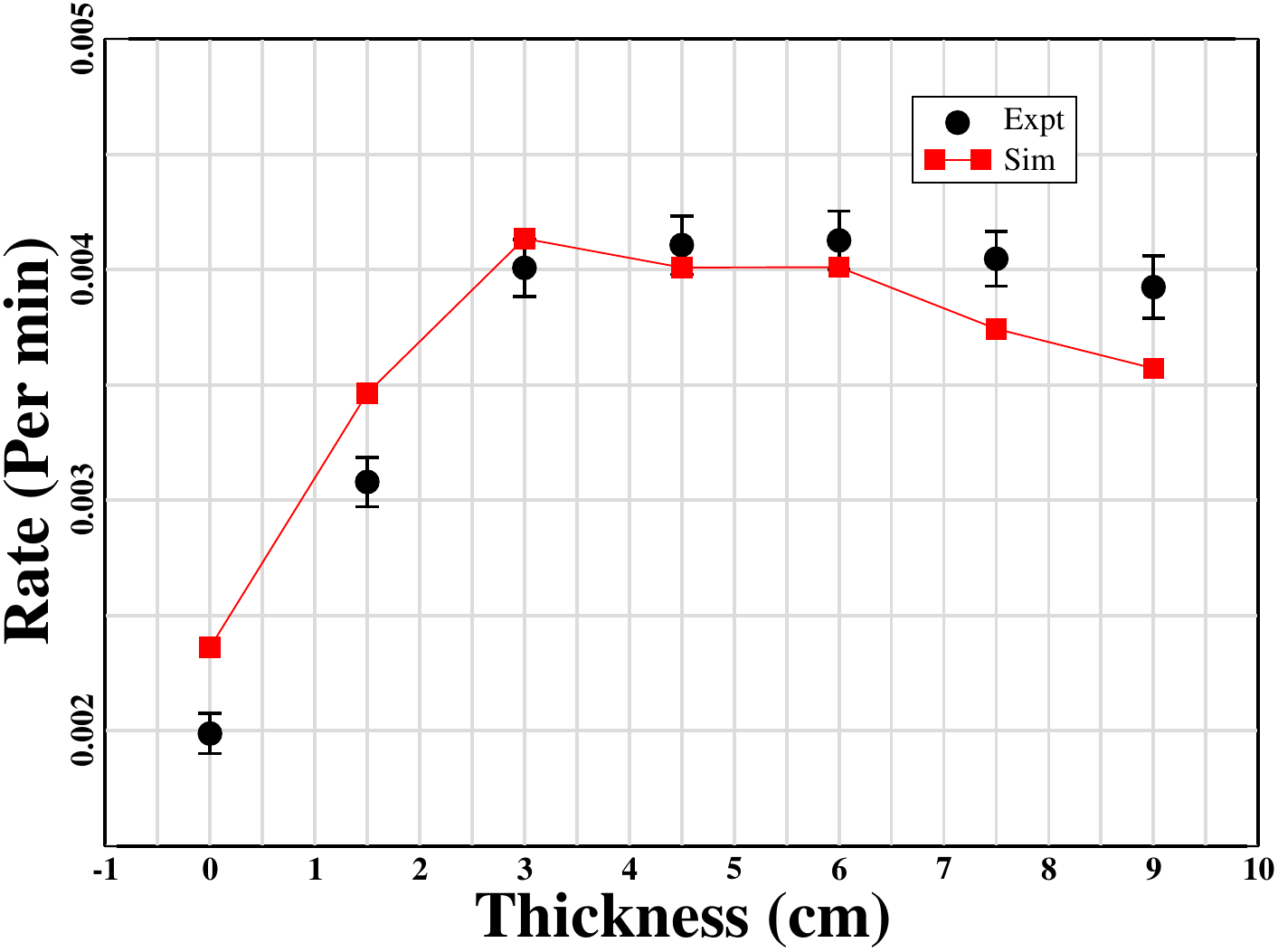}
        \caption{Comparison of experimental and simulation results.}
        \label{fig:simulation_experiment_plot}
    \end{subfigure}
    \hfill
    \begin{subfigure}[b]{0.7\textwidth}
        \centering
        \begin{subfigure}[b]{0.49\textwidth}
            \includegraphics[width=\textwidth]{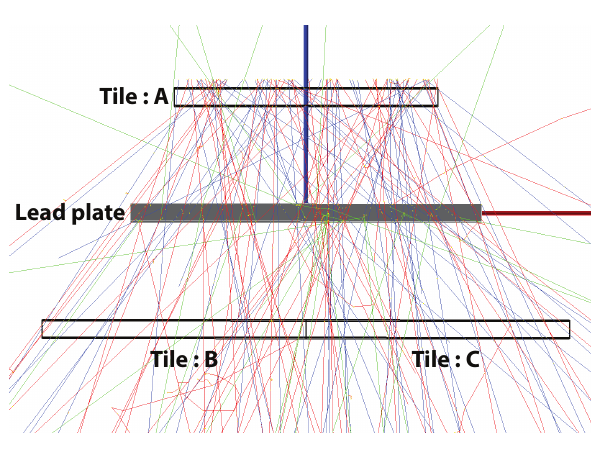}
        \end{subfigure}
        \hfill
        \begin{subfigure}[b]{0.49\textwidth}
            \includegraphics[width=\textwidth]{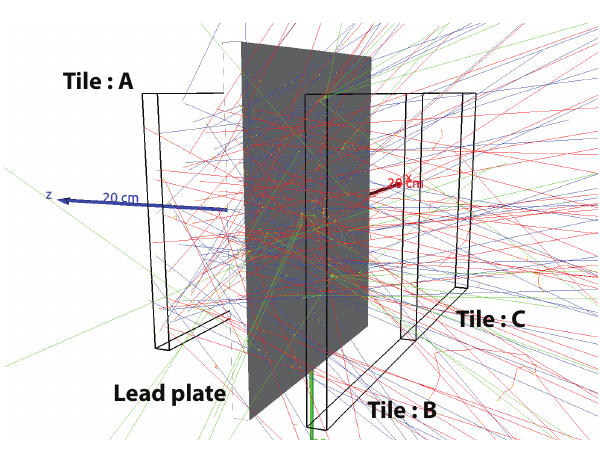}
        \end{subfigure}
        \caption{Geant4 simulation of the triangular detector setup.}
        \label{fig: geant4_geometry}
    \end{subfigure}
    \caption{Experiment and simulation conducted for secondary particle detection. (a) Comparison of experimental data with simulation, showing agreement in secondary shower behavior. (b) Geant4 model of the triangular configuration, detailing detector and absorber placement.}
\end{figure}

In this section, we describe a detector setup modeled in Geant4, optimized to capture secondary particle showers produced when muons interact with a lead absorber. The configuration consists of three detector plates arranged in a triangular layout: two plates (A and B) are placed parallel to each other beneath a lead plate, while a third plate (C) is positioned above the lead. This arrangement facilitates the detection of secondary particles, including electrons and gamma rays, generated as muons traverse the lead layer. The experimental setup, shown in Fig. \ref{fig: geant4_geometry} used a lead absorber of varying thicknesses (ranging from 1 cm to 6 cm) to analyze the secondary particle yield.

As discussed in \cite{ras2020}, the formation of the Rossi curve in such setups arises due to an initial increase in secondary particle production with absorber thickness, followed by a decline as the absorber captures a greater fraction of secondary particles. This expected trend is observed in our experimental data, confirming the effectiveness of the setup in capturing secondary showers with high accuracy.

A Geant4 simulation was carried out (Fig. \ref{fig: geant4_geometry}) with a muon flux generated using the CRY library, considering an initial muon energy spectrum corresponding to sea-level conditions.The physics list used  for this simulation was FTFP\_BERT \cite{Agostinelli2003}. 

The results, illustrated in Figure \ref{fig:simulation_experiment_plot}, demonstrate a good agreement between experimental and simulated data. As the lead thickness increases, the detection rate of secondary particles initially rises due to enhanced secondary shower production. However, beyond a critical thickness, absorption dominates, leading to a reduction in detected secondary particles.

\subsection{Muon absorption study}
\label{section: absorption}

\begin{figure}[h]
    \centering
    \includegraphics[width=1\textwidth]{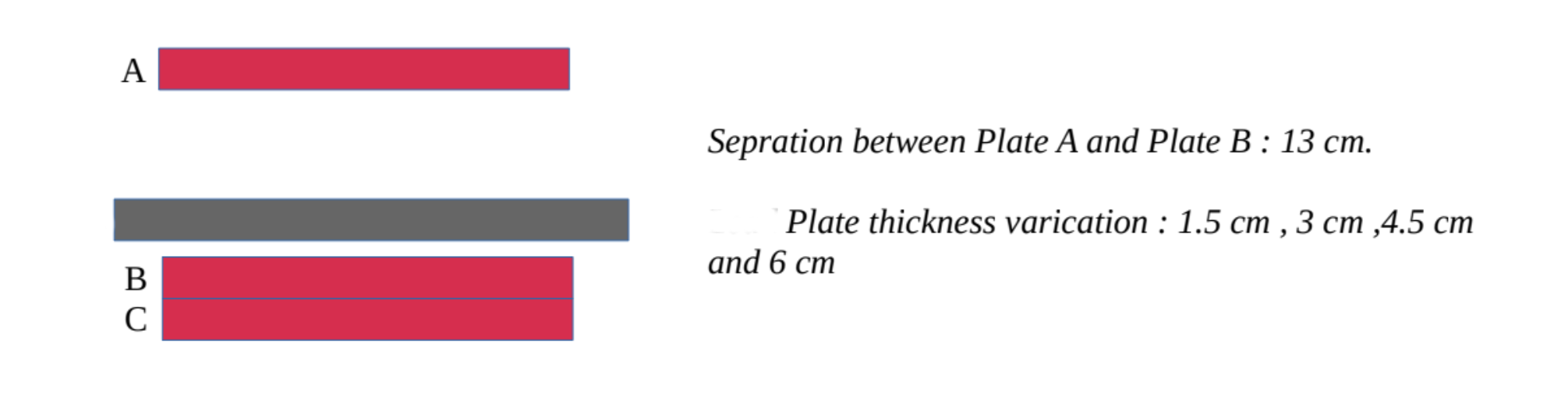}
    \caption{Diagram of the detector setup geometry with measurements. The labeled plates (A, B, and C) are separated at fixed distances to ensure consistent detection.}
    \label{fig: detector_setup}
\end{figure}

This experiment focuses on measuring muon absorption in stainless steel, aluminum, and concrete using the Cosmic Hunter setup. A corresponding numerical study was also carried out that estimated the amount of muon absorption due to the same materials with varying thicknesses. The experimental setup consisted of three detector plates labeled A, B, and C, arranged vertically, as shown in Fig.\ref{fig: detector_setup}. Plates A and B were placed 13 cm apart, while plate C was positioned immediately after plate B, without any gap between them. Muons passing through this setup were detected in coincidence mode.

Absorption studies were conducted with material plates of stainless steel, aluminum, and concrete. The former two were available in four thicknesses: 1.5 cm, 3 cm, 4.5 cm, and 6 cm, while the last material was available in three thicknesses: 2 cm, 4 cm and 6 cm. The goal was to quantify the attenuation of muons as they traversed different materials of increasing thickness. Before introducing the plates, background muon flux measurements were recorded for calibration. This ensured that subsequent variations in count rates were due to absorption and not fluctuations in environmental conditions.

\begin{figure}[h]
    \centering
    \begin{minipage}[b]{0.4\textwidth}
        \includegraphics[width=\textwidth]{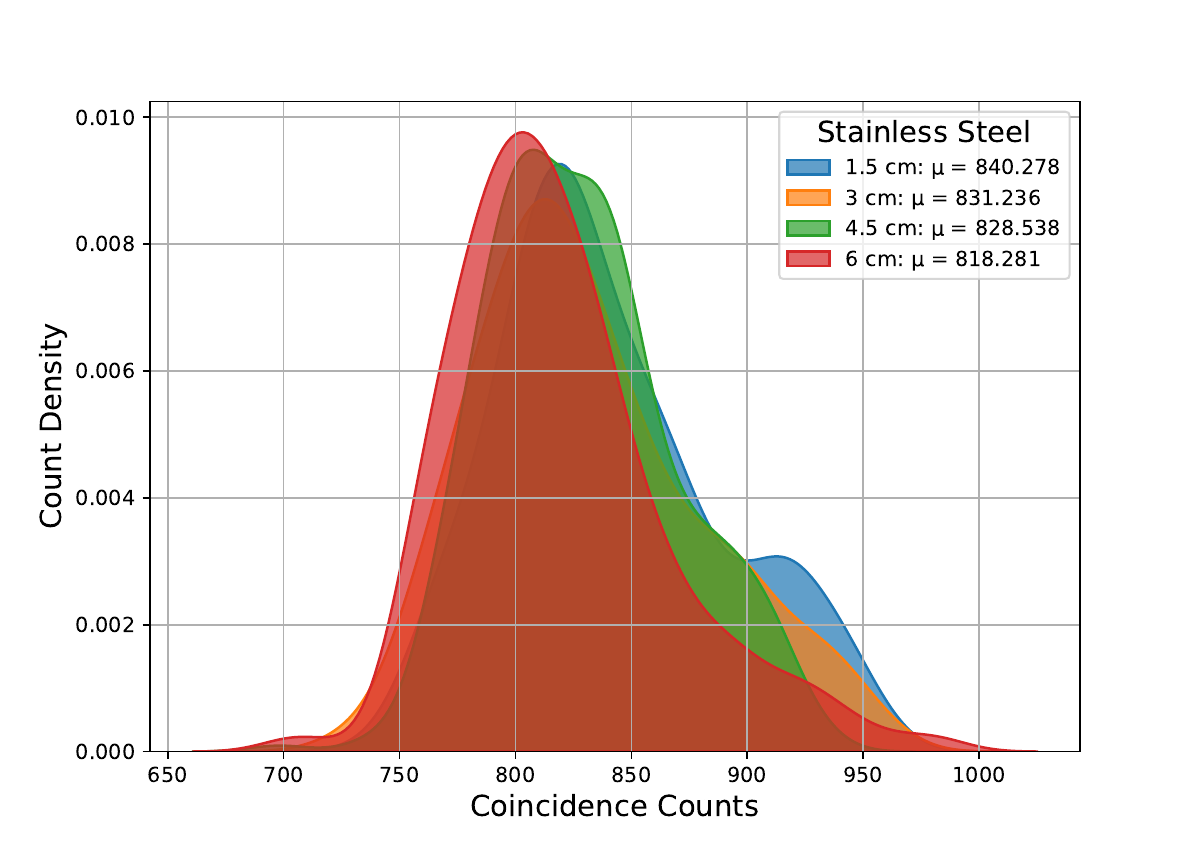}
        \caption{Ridge plot for different stainless steel plate thicknesses.}
        \label{fig: ridg_steel}
    \end{minipage}
    \hspace{0.01\textwidth}
    \begin{minipage}[b]{0.4\textwidth}
        \includegraphics[width=\textwidth]{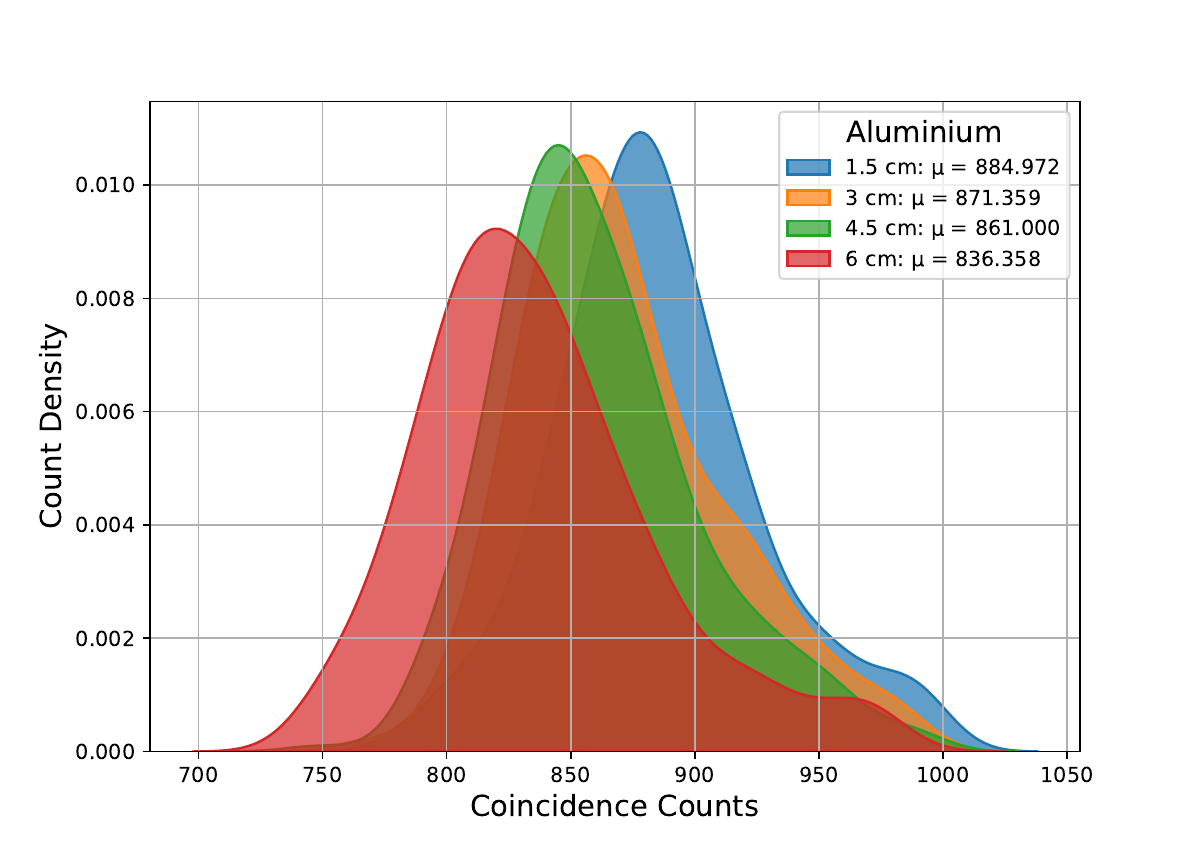}
        \caption{Ridge plot for different aluminum plate thicknesses.}
        \label{fig: ridg_aluminum}
    \end{minipage}
    \\
    \begin{minipage}[b]{0.4\textwidth}
    	\includegraphics[width=\textwidth]{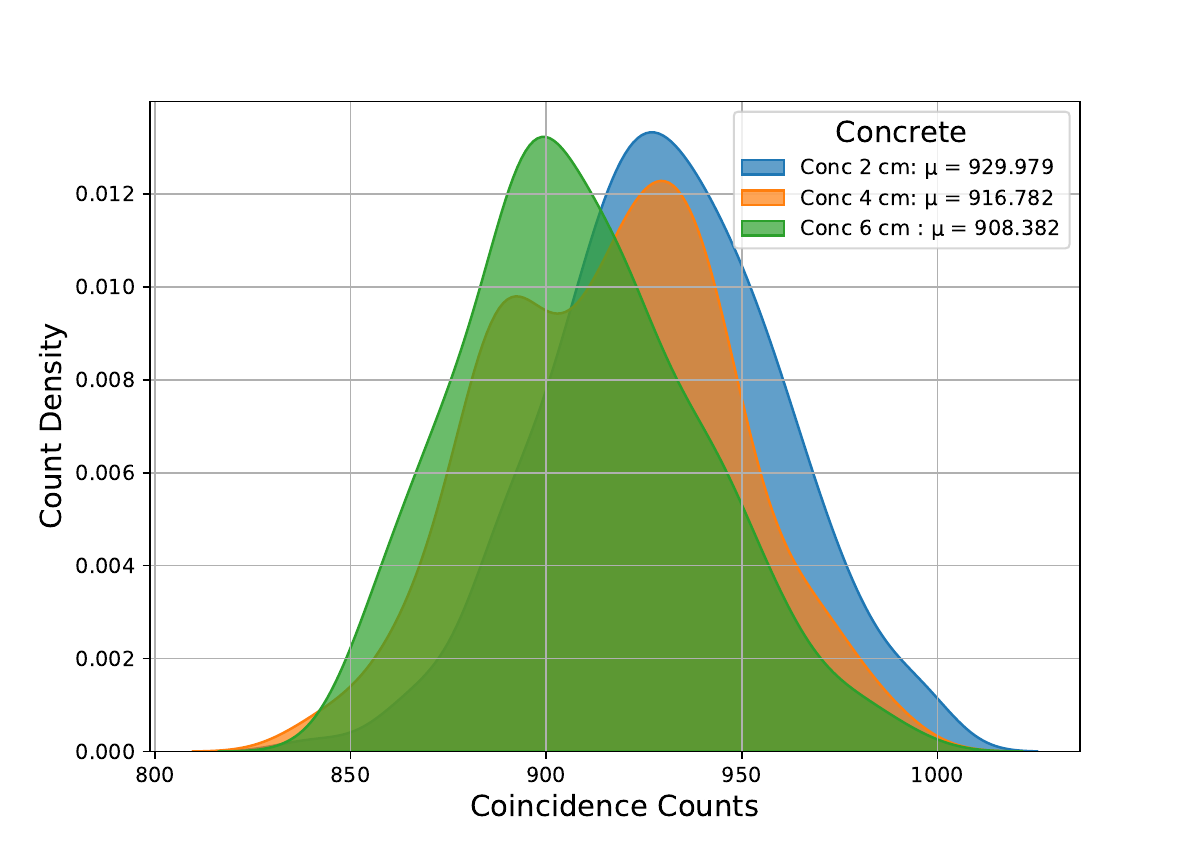}
    	\caption{Ridge plot for different concrete plate thicknesses.}
    	\label{fig: ridg_conc}
    \end{minipage}
\end{figure}

For each material and thickness, data was collected over a full week with hourly recordings. The muon counts were analyzed through ridge plots to determine the mean absorption values. Figures~\ref{fig: ridg_steel}, \ref{fig: ridg_aluminum}, and \ref{fig: ridg_conc}, display the ridge plots corresponding to stainless steel, aluminum, and concrete respectively.

\begin{figure}[h]
    \centering
    \begin{minipage}[b]{0.45\textwidth}
        \includegraphics[width=\textwidth]{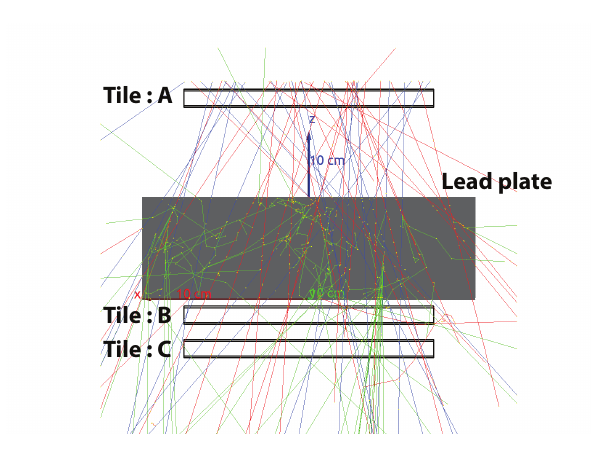}
    \end{minipage}
    \hspace{0.05\textwidth}
    \begin{minipage}[b]{0.45\textwidth}
        \includegraphics[width=\textwidth]{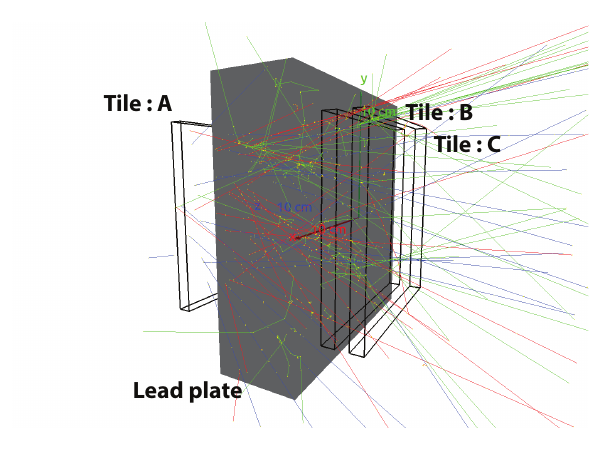}
    \end{minipage}
    \caption{Geant4 simulation of the detector and material plate.}
    \label{fig:geant4_simulation}
\end{figure}

A Geant4-based simulation was used to compare with experimental data. The simulation incorporated the measured muon flux inside the room (\(0.5 \, \text{muons/cm}^2/\text{min}\)) and recreated the detector stack with identical dimensions and material configurations. The simulated setup is illustrated in Figure~\ref{fig:geant4_simulation}.

\begin{figure}[h]
	\centering
	\begin{minipage}[b]{0.45\textwidth}
		\includegraphics[width=\textwidth]{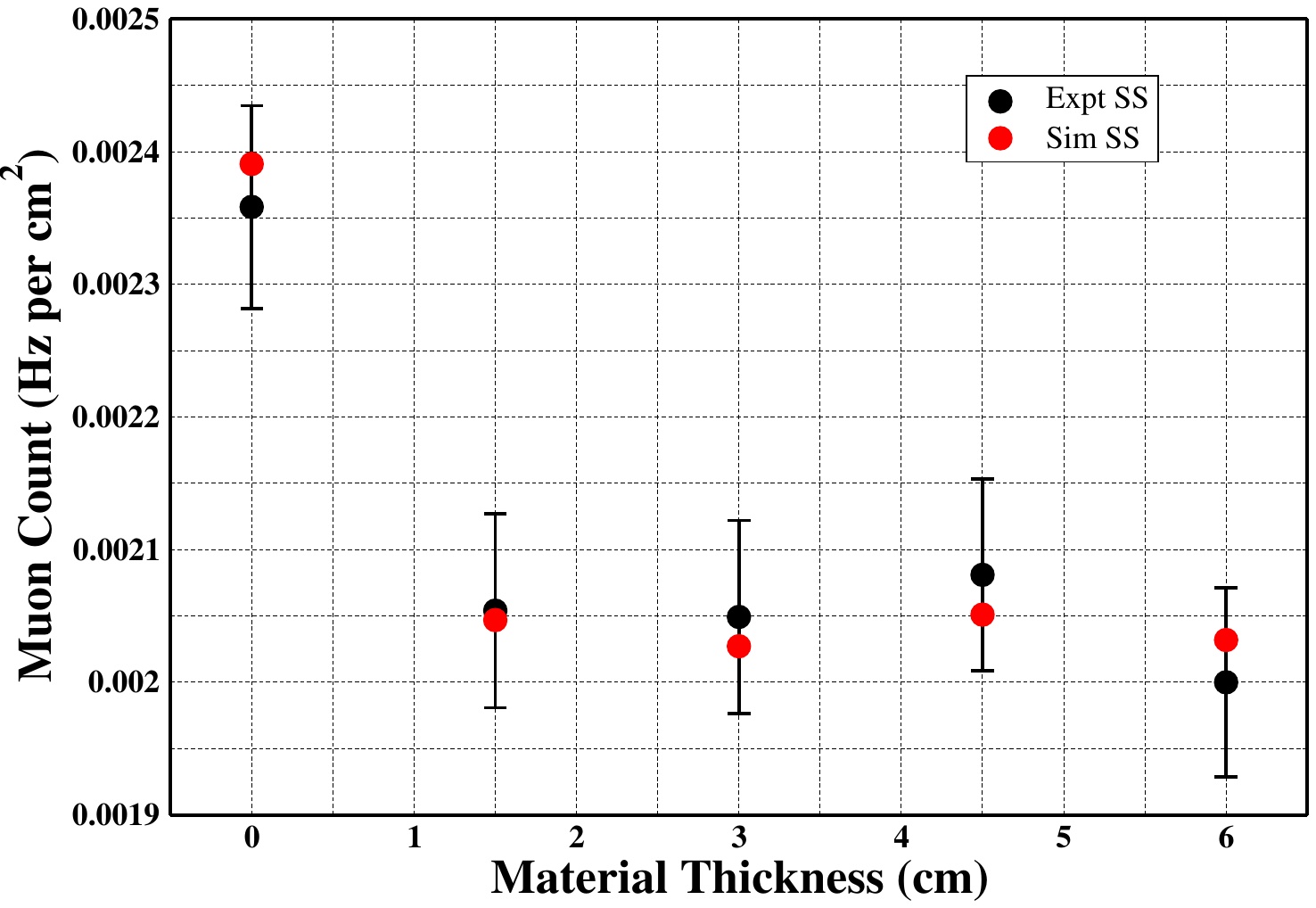}
		\caption*{(a) Stainless Steel}
	\end{minipage}
	\hspace{0.05\textwidth}
	\begin{minipage}[b]{0.45\textwidth}
		\includegraphics[width=\textwidth]{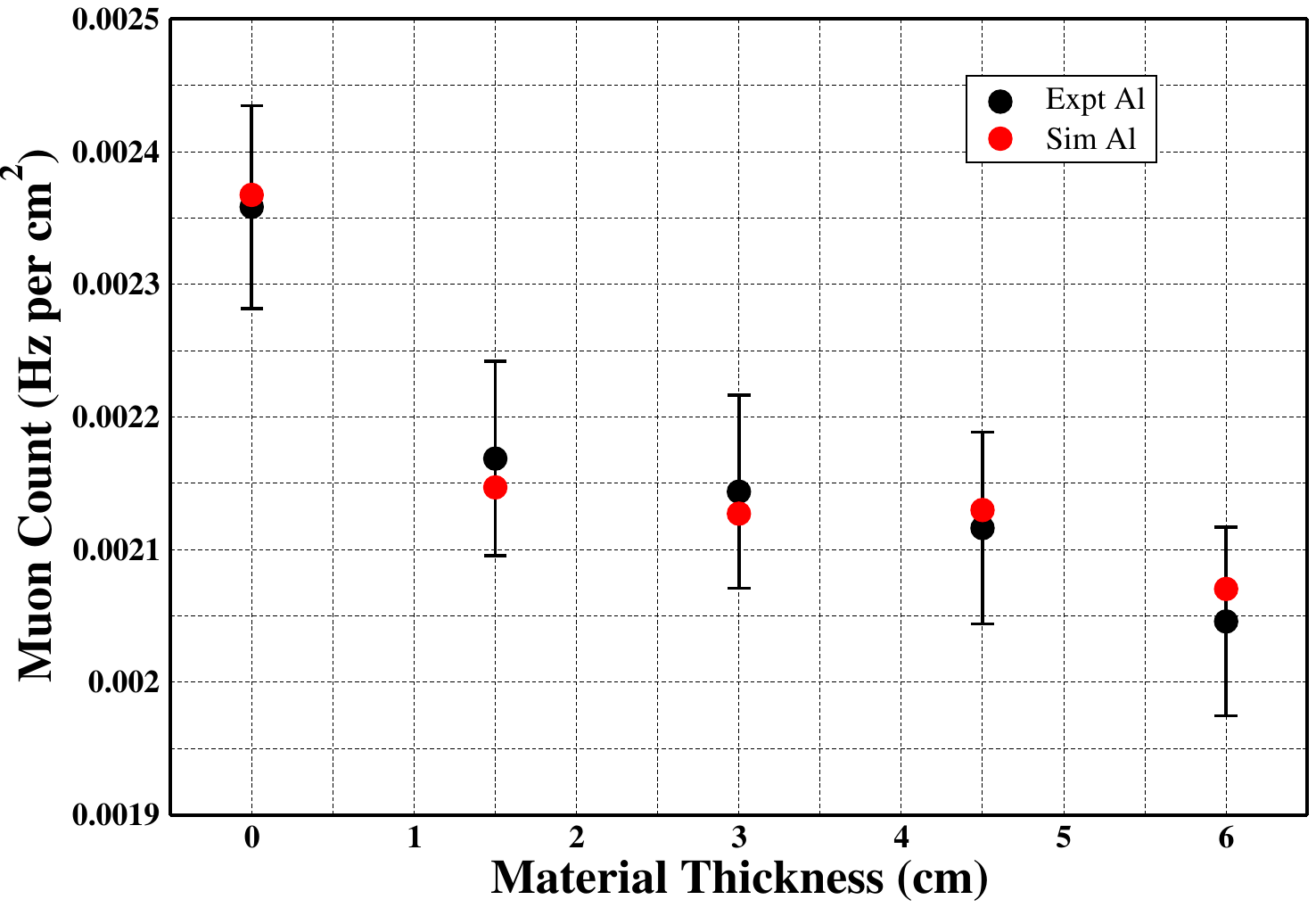}
		\caption*{(b) Aluminum}
	\end{minipage}
	\\
	\begin{minipage}[b]{0.45\textwidth}
		\includegraphics[width=\textwidth]{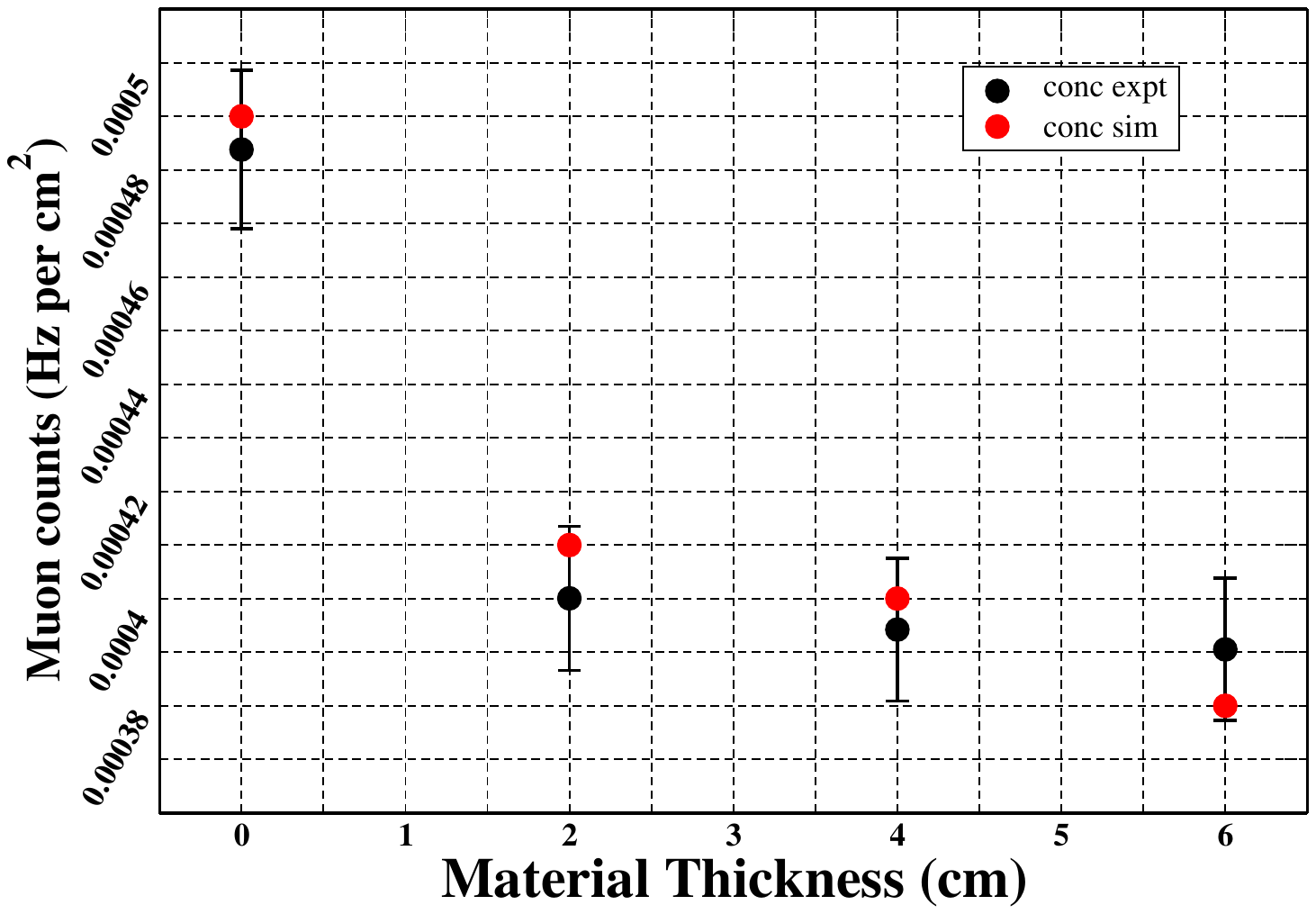}
		\caption*{(c) Concrete}
	\end{minipage}
	\caption{Comparison of experimental and simulation results for muon absorption in (a) stainless steel, (b) aluminum, and (c) concrete. The experimental data (markers) align well with the Geant4 simulation (solid lines), demonstrating the accuracy of the absorption measurements.}
	\label{fig: comparison_plots}
\end{figure}

The experimental and simulated results, including muon count distributions and spatial absorption patterns, are shown in Fig.\ref{fig: comparison_plots}. This agreement between the experimental and simulated data further confirms that the muon detection system can effectively capture variations in muon absorption across different materials.

\section{MARS (Muon Absorption in Rigid Structures)}
\label{section: MARS}

This final part of our study investigates the absorption of muons in rigid structures, focusing on how muons interact with relatively complex systems composed of standard construction materials such as concrete blocks. This study aims to assess the feasibility of using MAT based on the setup described in this work to detect voids or irregularities within materials, enabling structural analysis and imaging.

\begin{figure}[h]
    \centering
    \begin{minipage}[b]{0.42\textwidth}
    	\centering
    	\includegraphics[width=\textwidth]{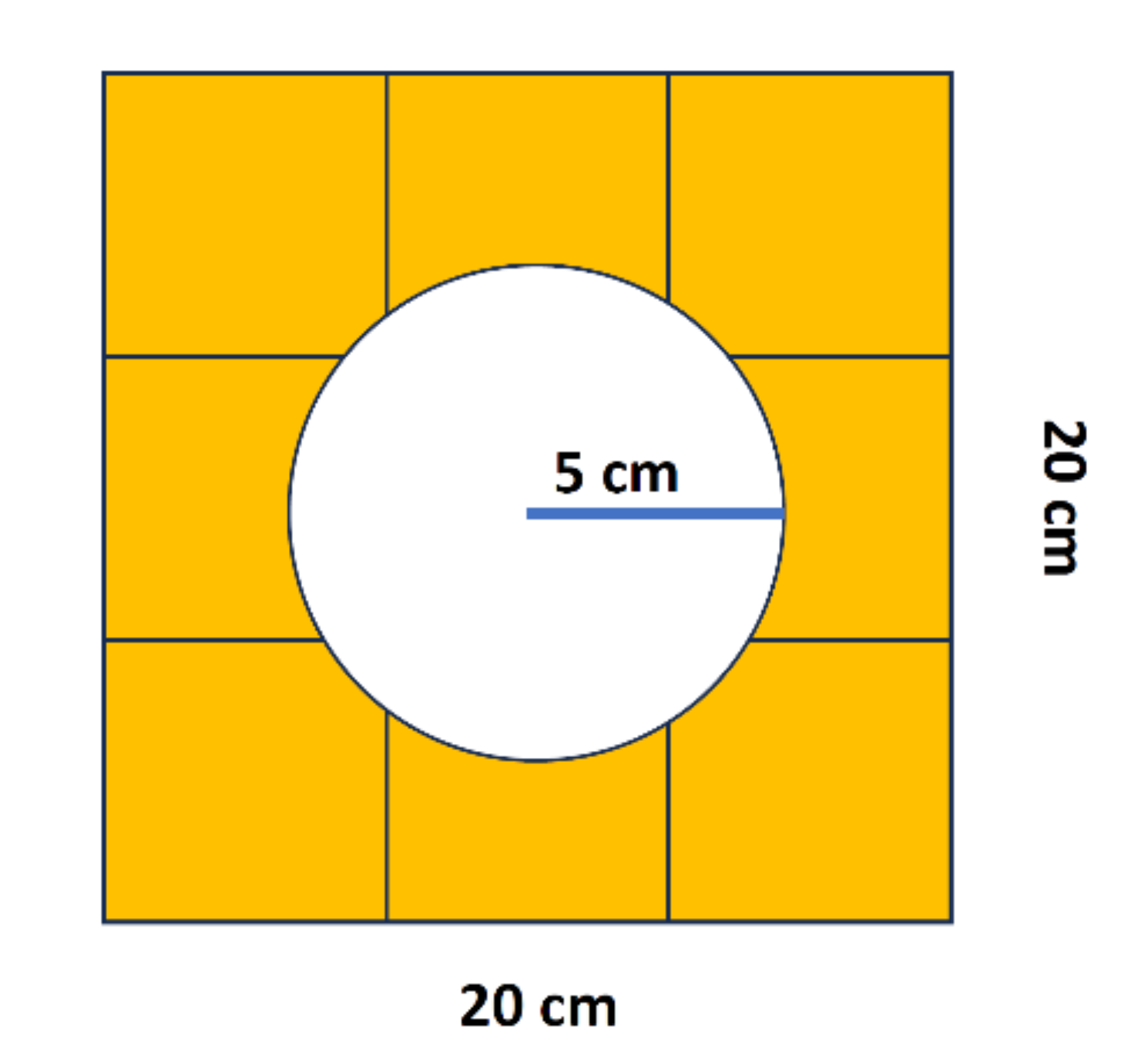}
    	\caption{Concrete block with a central hole used for muon absorption studies.}
    	\label{fig: concrete_with_hole}
    \end{minipage}
    \hspace{0.05\textwidth}
    \begin{minipage}[b]{0.45\textwidth}
        \centering
        \includegraphics[width=\textwidth]{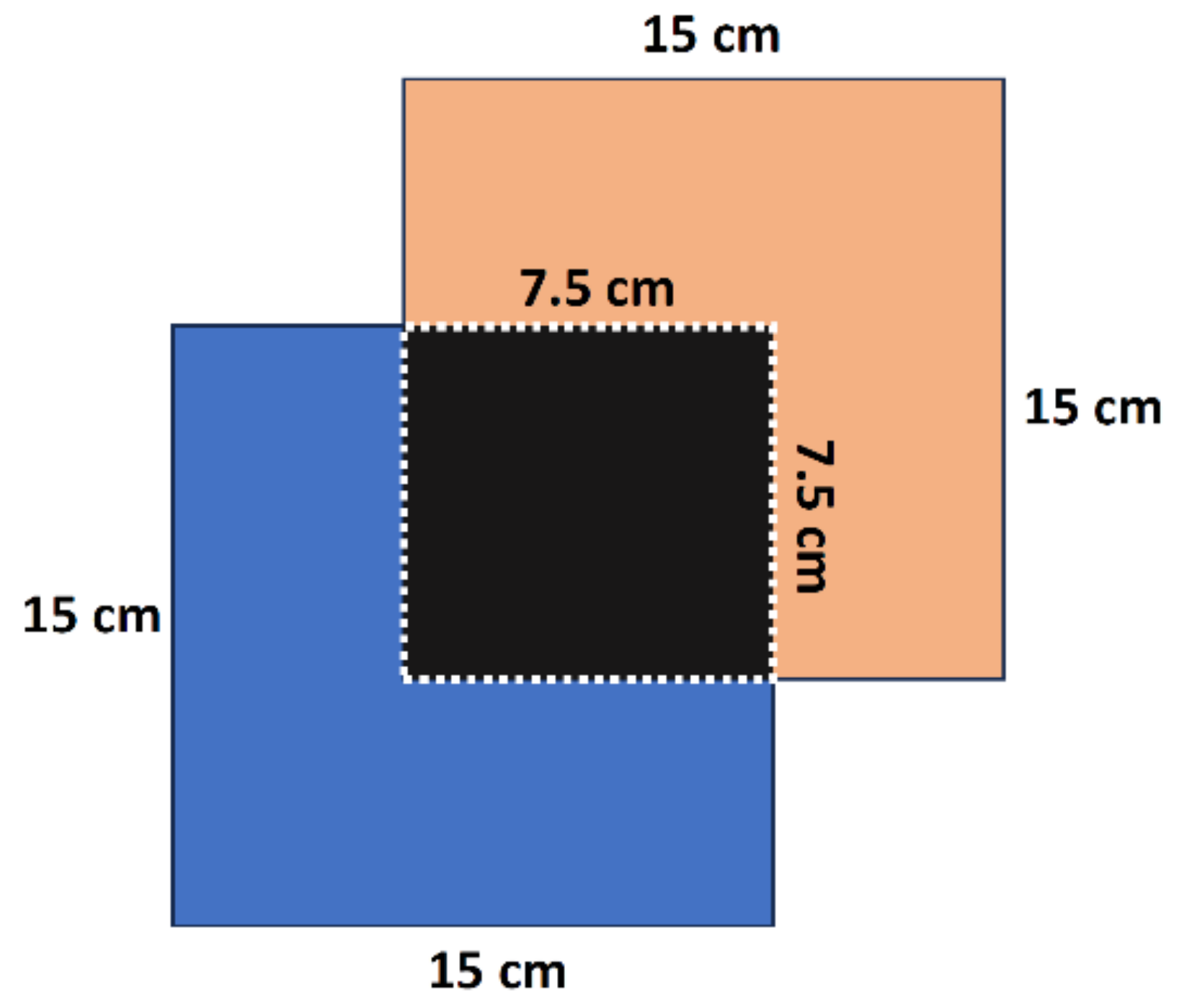}
        \caption{Overlapped detector plate configuration with labeled measurement regions.}
        \label{fig: overlapped_tiles}
    \end{minipage}
\end{figure}

The target material for this study was a concrete slab measuring \(20 \times 20 \times 2\) cm\(^3\), featuring a centrally located void with a 5 cm radius, as shown in \ref{fig: concrete_with_hole}. The thickness of the concrete structure was varied from 2cm to 6cm. The presence of this void provided a well-defined test scenario for evaluating the detector’s capability to identify structural inhomogeneity. For this purpose, the three scintillating tiles, each measuring \(15 \times 15 \times 1\) cm\(^3\), were arranged in an overlapping grid pattern to enhance spatial resolution, as shown in \ref{fig: overlapped_tiles}. This layout, in which the effective detector layout was divided into a \(3 \times 3\) grid of small square regions (indicated in \ref{fig: concrete_with_hole}), allowed multiple tiles to detect muons only over the overlapped tile area, thereby enabling improved localization of absorption effects within the target material.

\begin{figure}[h]
    \centering
    \begin{subfigure}[b]{0.45\textwidth}
        \includegraphics[width=\textwidth]{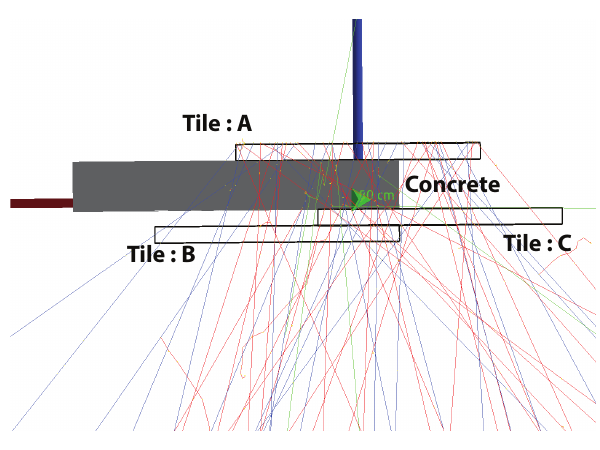}
        \label{fig:G4_front}
    \end{subfigure}
    \hspace{0.05\textwidth}
    \begin{subfigure}[b]{0.45\textwidth}
        \includegraphics[width=\textwidth]{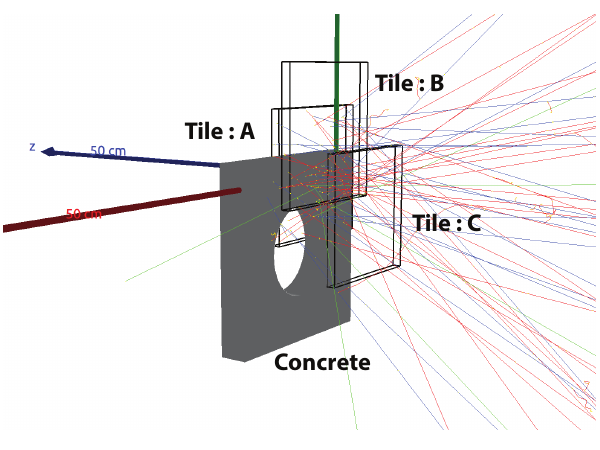}
        \label{fig:G4_side}
    \end{subfigure}
    \caption{Geant4 simulation for overlapped detector setup.}
    \label{fig:geant4_simulation_overlapped}
\end{figure}

To validate the experimentally measured data, a Geant4 simulation of the detector and material configuration was performed once again. The simulation involved generating 50,000 cosmic ray muons and the use of FTFP\_BERT physics model. Figure~\ref{fig:geant4_simulation_overlapped} presents different views of the simulated detector system, showing close resemblance with the experimental setup. The resulting simulation provided a basis for comparison with experimental data, ensuring that experimentally detected muon absorption variations corresponded to actual material properties rather than random fluctuations.

\begin{figure}[h]
	\centering
	\begin{minipage}[b]{0.3\textwidth}
		\includegraphics[width=\textwidth]{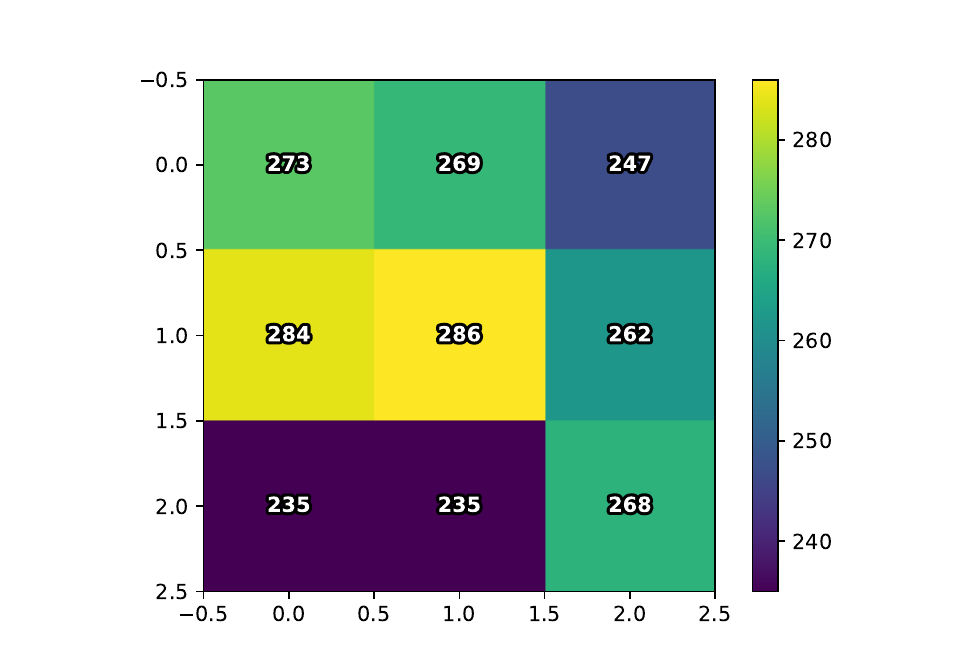}
		\caption*{(a) Experimental 2 cm}
	\end{minipage}
	\hspace{0.05\textwidth}
	\begin{minipage}[b]{0.3\textwidth}
		\includegraphics[width=\textwidth]{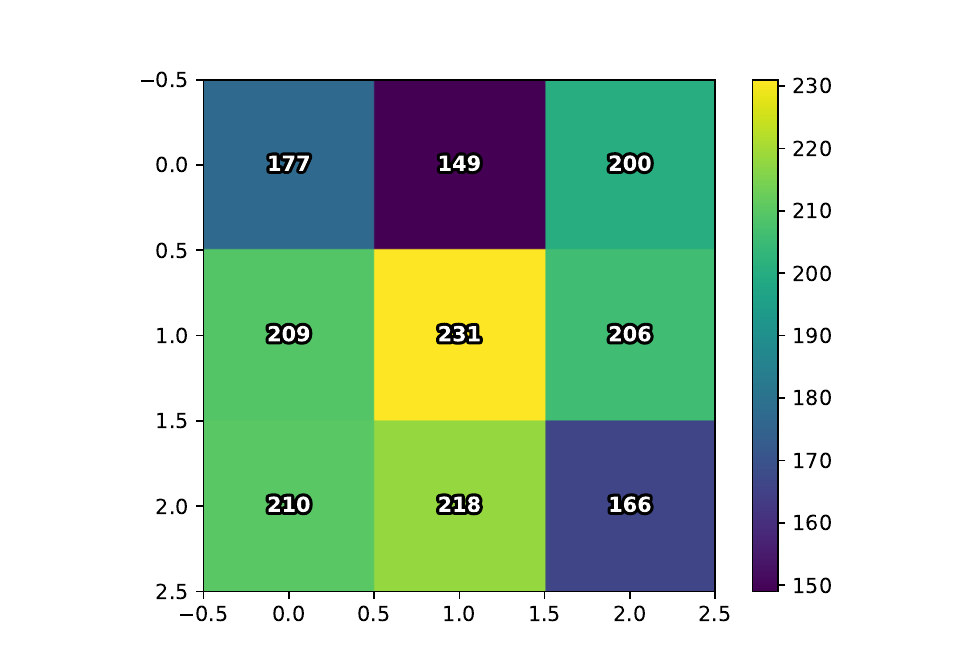}
		\caption*{(b) Experimental 4 cm}
	\end{minipage}
	\hspace{0.05\textwidth}
	\begin{minipage}[b]{0.3\textwidth}
		\includegraphics[width=\textwidth]{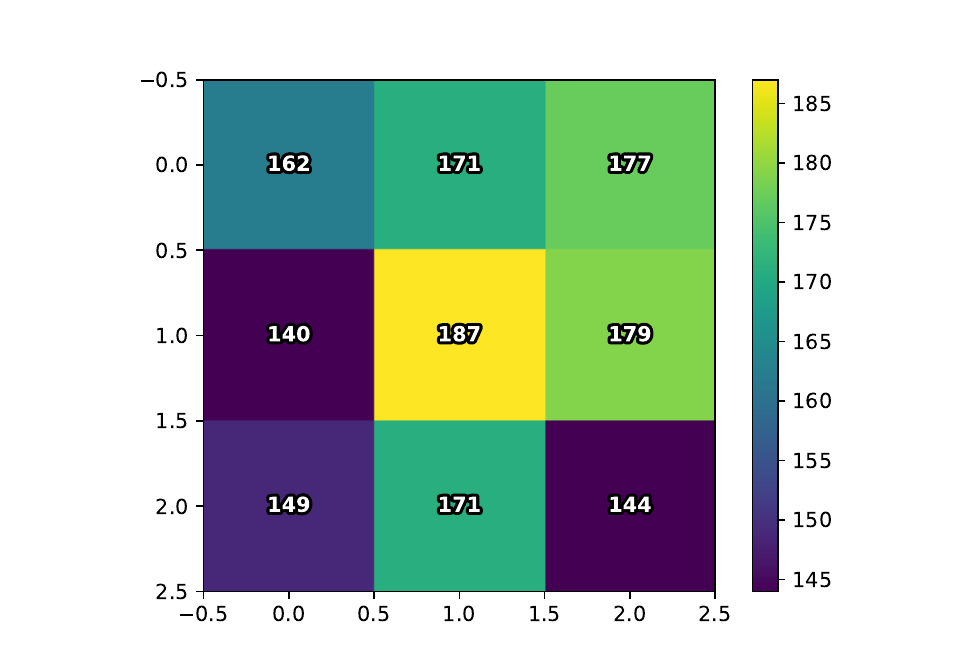}
		\caption*{(c) Experimental 6 cm}
	\end{minipage}
	
	\vspace{0.3cm} 
	\begin{minipage}[b]{0.3\textwidth}
		\includegraphics[width=\textwidth]{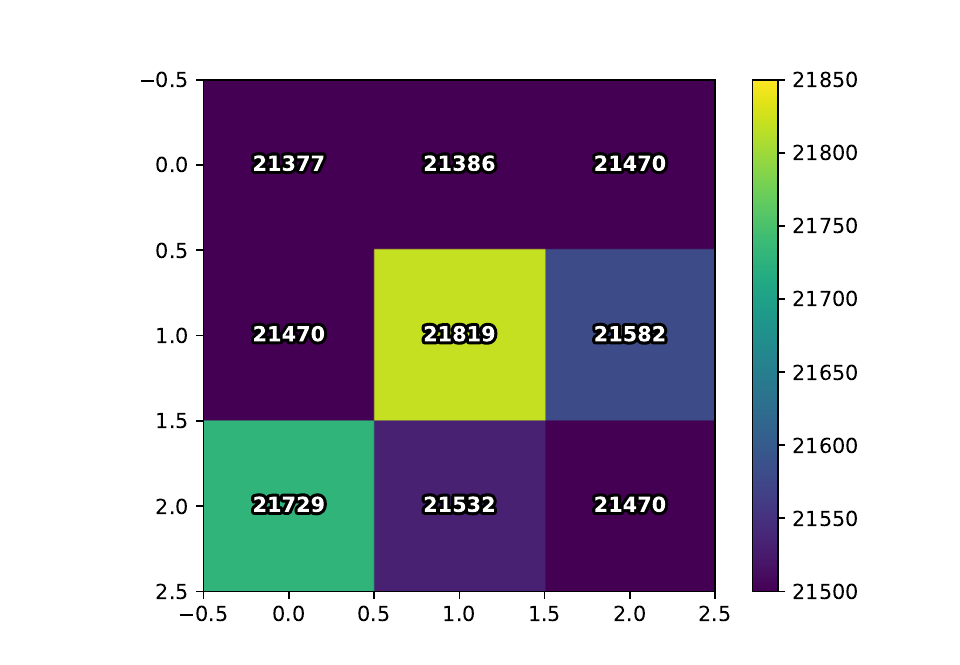}
		\caption*{(d) Simulation 2 cm}
	\end{minipage}
	\hspace{0.05\textwidth}
	\begin{minipage}[b]{0.3\textwidth}
		\includegraphics[width=\textwidth]{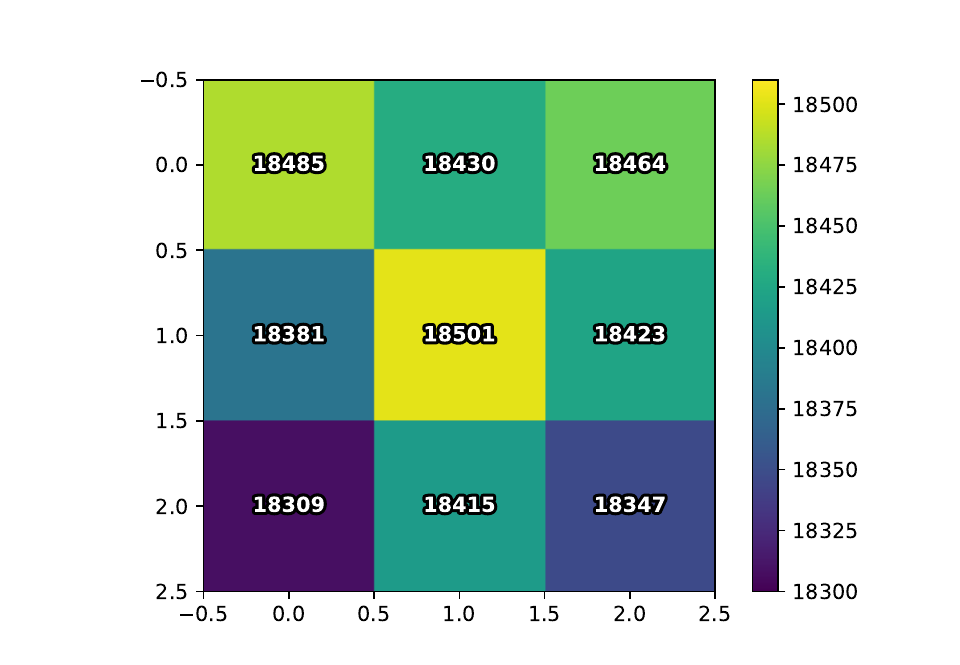}
		\caption*{(e) Simulation 4 cm}
	\end{minipage}
	\hspace{0.05\textwidth}
	\begin{minipage}[b]{0.3\textwidth}
		\includegraphics[width=\textwidth]{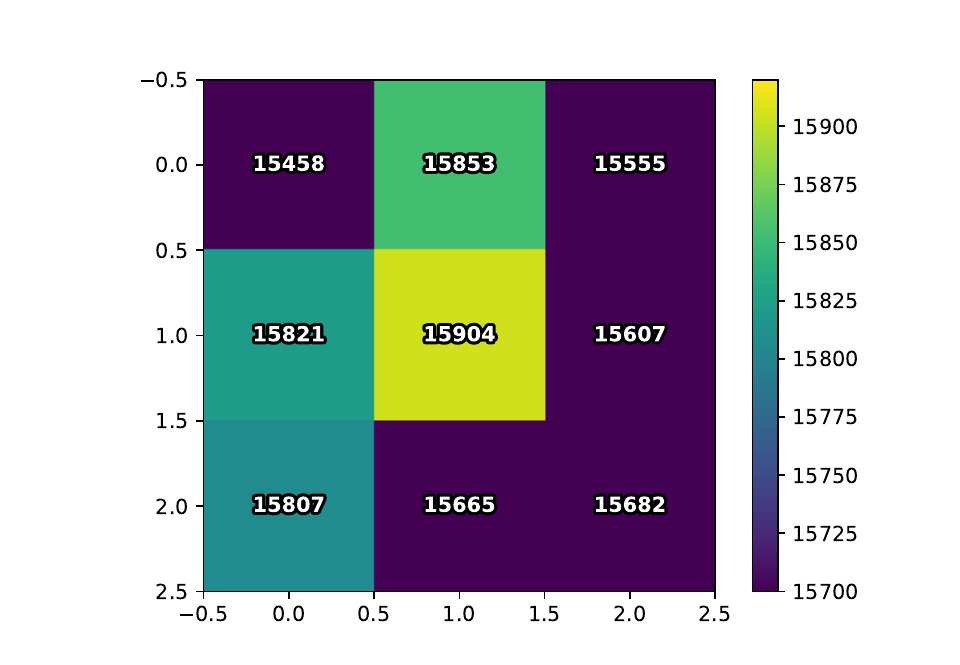}
		\caption*{(f) Simulation 6 cm}
	\end{minipage}
	
	\caption{Comparison of experimental and simulation results for the overlapped detector setup with 2 cm, 4 cm, and 6 cm thick concrete plates. Both experimental and simulated data confirm the presence of a central void due to higher muon counts in that region.}
	\label{fig:pixel_detector_results}
\end{figure}

\begin{table}[h]
	\centering
	\caption{Difference in muon counts between the central and surrounding regions for both experimental and simulated data.}
	\begin{tabular}{|c|c|c|c|}
		\hline
		\multicolumn{4}{|c|}{\textbf{Experimental Data}} \\
		\hline
		Thickness & Avg Surrounding Counts & Central Counts & \% Difference \\
		\hline
		2 cm & 259.125 & 288 & 10.03 \% \\
		4 cm & 191.875 & 231 & 16.94 \% \\
		6 cm & 161.625 & 187 & 13.57 \% \\
		\hline
		\multicolumn{4}{|c|}{\textbf{Simulation Data}} \\
		\hline
		Thickness & Avg Surrounding Counts & Central Counts & \% Difference \\
		\hline
		2 cm & 21523 & 21819 & 1.3 \% \\
		4 cm & 18406 & 18501 & 0.5 \% \\
		6 cm & 15681 & 15904 & 1.4 \% \\
		\hline
	\end{tabular}
	\label{table:pixel_experiment_results}
\end{table}

The results, shown in Figure~\ref{fig:pixel_detector_results}, confirm that the overlapped detector configuration successfully identified a void in the center of the concrete slabs, as indicated by the higher muon coincidence counts in the void region. Similarly, Table~\ref{table:pixel_experiment_results} presents a quantitative comparison of the average muon counts in the central void and the surrounding areas for both experimental and simulated data. The higher muon counts observed in the central region align with the expected reduction in material density. While it is true that there are fluctuations in the percentage of increase in the number of muons through the hole, there is no doubt that the number is always more than that through the concrete in the surroundings.

\section{Conclusion and future plans}
\label{section: conclusion}

In this study, we have investigated the possibility of using MAT for monitoring archaeological, civil and industrial structures. The final goal is to develop a hybrid muography system that uses both MAT and MST to image the internals of an intermediate-sized structure. By intermediate-sized structures, we have meant strucutures of sizes that are small enough to place muon detectors on either side of the object and large enough to cause detectable muon absorption.

For the present work, we have used a very easily available educational tool procured from CAEN (the cosmic hunter). The tool itself has been validated and calibrated by several basic studies related to cosmic ray muons. It may be noted however that it is not essential to use only this product, or even another commercial one. Through dedicated effort, an equivalent system can also be developed in a laboratory. The major concern related to any such new system will be its efficiency and ability to detect muon absorption which can be established by carrying out basic experiments as done during this project. We have tried to validate experimental measurements as much as possible, using Geant4 simulations using FTFP\_BERT physics model. For this purpose, the CRY library has been used as the cosmic ray event generator.

The efficiency of each scintillation tile of the cosmic hunter has proved to be very close to 90\%, which satisfied the requirements of the proposed goal reasonably well. Experiments were also carried out to check the possibility of detecting secondary particles generated due to cosmic rays interacting with intervening materials. This relatively difficult experiment has yielded results that are in good agreement with previously reported values and numerical simulation carried out during the present work. This fact has increased our confidence in the experimental setup.  

Work related to muon absorption has been initiated by carrying out a comparison between the experimental measurements and the simulated values for absorption through three materials, namely, stainless steel, aluminum, and concrete, all three being standard building materials. The experimentally obtained muon absorption values have been found to match closely with the simulated results. Finally, the ability of the setup to detect muon absorption was tested for concrete blocks of varying thickness having a hole in its center. For this purpose, scintillation tiles were used in partially overlapped geometries in order to enhance localization abilities of the system. The resulting setup was found to be quite successful in locating the void within the concrete block, despite unusual modifications in the placement of scintillation tiles. 

The findings of the present investigation indicate that MAT based on simple instrumentation can be profitably used to visualize internal features of structures, having utility in non-destructive testing, structural health monitoring, and archaeological exploration of objects of intermediate size. The process is inherently time-consuming, as achieving a statistically significant muon count requires prolonged data acquisition. The precision of the system is further constrained by detector efficiency, secondary generation, etc. In future, we plan to improve detector sensitivity, and optimize data processing algorithms and flow. We also hope to extend this investigation by adding MST to this setup. It is expected that combination of both absorption and scattering data will significantly improve monitoring abilities of the resulting setup.

\acknowledgments
The authors acknowledge the help and support from their respective Institutes and Universities. This research was supported in part by the SERB research grant SRG/2022/000531. Authors Mr. P.Pallav and Dr. P. Bhattacharya acknowledge the Science and Engineering Research Board and SRG scheme for the necessary support.

\bibliographystyle{JHEP}
\bibliography{PaperBiblio}

\end{document}